# Influence of particle size on ice nucleation and growth during the freeze-casting process

Sylvain Deville[1], Eric Maire[2], Audrey Lasalle[1], Agnes Bogner[2], Catherine Gauthier[2], Jérôme Leloup[1], Christian Guizard[1]

[1] Laboratoire de Synthèse et Fonctionnalisation des Céramiques, UMR3080 CNRS/Saint-Gobain CREE, 550 Avenue Alphonse Jauffret, BP 20224, 84306 Cavaillon Cedex, France

[2] Université de Lyon, INSA-Lyon, MATEIS CNRS UMR5510, 7 avenue Jean Capelle, F-69621 Villeurbanne, France

**Abstract**

The solidification behaviour of suspensions of alumina particles during directional solidification is investigated here by in situ observations using X-ray radiography and tomography. The objective of this study is to assess the influence of particle size on the solidification behaviour of the suspensions during the early stages of solidification. Four powders with particle size in the range 0.2–3.4 µm (median size) were investigated. Solidification is obtained by cooling at a constant rate, starting from room temperature. Attention is specifically paid to the nucleation and growth behaviour of the ice crystals in these suspensions. We propose that the nucleation of ice crystals is controlled by the particle size, the surface of the particles acting as nucleation sites. Smaller particle size leads to a lower degree of supercooling since nucleation and growth can proceed at higher temperature than with larger particles. The initial interface velocity is dependent on the degree of supercooling, and controls the extent of the initial structural gradient in the resulting porous materials.

## 1. Introduction

The templating of porous materials using ice, through the process of freeze-casting, have seen a greatly increased attention during the past few years, regarding not only ceramics, but also polymers[1] and metals[2]. The process is environmentally-friendly, using water as a removable template, highly versatile, and the resulting structures highly tuneable by various tweaks of the process such as an improved control of the nucleation conditions or the use of additives affecting the morphologies of the growing crystals[3]. The typical processing conditions include directional solidification, using a cooling step starting at room temperature. Under these conditions, nucleation and growth in the suspension has to occur before the ice crystals can reach a steady state, progressively growing in the temperature gradient, and yielding homogeneous and directional materials, after sublimation of the ice and sintering of the resulting green body. In all cases, this initial nucleation and growth stage results in the presence of a structural gradient close to the cooling surface, corresponding to the progressive selection of the stable crystals structure[4]. Such gradient has been reported numerous times, such as figure 2 of reference [5], figure 12 of reference [6], figure 8 of reference [7], figure 6a-b of reference [8], figure 2 of reference [9], and a number of other studies. The presence of this structural gradient is explained by the initial conditions during the freezing stage, and is characterized by two distinct regions. A first region, with no porosity, corresponds to the formation of amorphous ice during the very first stages of solidification[10], where supercooling effects are probably present. A second region, with a gradient of pore size, is related to the nucleation, growth, and selection of the preferred population of ice crystals[4]. Being related to the initial nucleation and growth properties, the gradient cannot be avoided so far, although it might be controlled to some extent. Applying an electric field will favour the electromigration towards or away from this initial transition[9], and can partially affect the extent of this structural gradient. We investigate here the influence of particle size on the characteristics of this initial transition and in particular on the nucleation behaviour in this system. X-Ray radiography and tomography experiments were used here to investigate the initial instants of solidification and the arrangement of the crystals and concentrated particles after complete solidification. The experiments were conducted at the ESRF, on the beamline ID19.

## 2. Experimental methods

All the experimental methods and powders characteristics have been described in reference [4] and [11]. All the results presented here have been obtained with a 5°C/min cooling ramp, starting from room temperature.

## 3. Results

The solidification interface velocity can be measured from the pictures sequence acquired by radiography. The measurements of interface velocities for the various powders are given in figure 1. The acquisition frequency (3Hz) is still too low to observe precisely the stage where the ice nuclei reach their critical size and keep growing irreversibly. In the time span (0.3 sec.) between the last frame with no crystals and the next one, crystals have already grown a few hundreds of micrometers; there is therefore an important uncertainty on the first point of the plot for each powder. The minimum growth velocity during these first instants can accordingly be estimated to be between 750 microns/s and 1200 microns/s, depending on the powder, and is probably faster than that. It is nevertheless clear that the initial growth velocity is related to the particle size, larger particles leading to greater initial interface velocity. Interestingly, all of the measurements fall on the same interface velocity vs. time trajectory, but with a different initial interface velocity. The arrows in figure 1 indicate the velocity of the interface when it reaches the top of the observation windows. Again, the larger the particle size, the greater the velocity at the end of the observation window and the smaller the structural periodicity in the frozen sample, which can clearly be observed on section 1 of figure 3.

Using the tomography scan performed on the frozen body, the solidified structure corresponding to the initial instants observed in figure 1 can be reconstructed. The corresponding cross-sections in the xz plane (parallel to the growth direction) and xy plane (perpendicular to the growth direction) are shown, respectively, in figure 2 and 3. The z location of the xy cross-sections of figure 3 are indicated in figure 2. In all cases, a morphological transition can be observed, with an intermediate stage where two populations of crystals can be observed, as reported previously[4]. The extent of this transition zone is dependent on the initial interface velocity, the higher the initial interface velocity, the larger the transition stage.

Measurements of the temperature close to the nucleation surface were performed at the same time, so that an estimation of the freezing temperature is available. We define here the freezing temperature as the time where the ice nuclei reach their critical size and keep growing irreversibly, so that they can actually be observed. The temperatures were not measured exactly at the surface, but the thermocouple is situated slightly below the nucleation surface. We can reasonably expect a small difference between the measured temperature and the true nucleation temperatures, but this should not affect the relative variations between the different powders, since the measurements were performed at the same location for all experiments. The freezing temperature is strongly dependent on the particle size, as shown in figure 4. Unfortunately, the freezing temperature for the largest particle size was not measured, due to difficulties with the experimental setup. It is nevertheless clear that the larger the particle size, the lower the freezing temperature. As a consequence, the lower the initial freezing temperature, the larger the initial interface velocities.

# 4. Discussion

All the experiments were performed by starting from the room temperature, and cooling at a constant rate of 5°C/min, until complete solidification was achieved. Ice crystals nuclei are therefore never present initially in the suspension and nucleation has to occur before growth can proceed and a steady state of growth can be achieved. We propose here that the particle size has a direct influence on the ice crystals nucleation in the suspension. The presence and role of supercooling was already mentioned[10] to explain the structure formation mechanisms when the suspensions are solidified by direct immersion in liquid nitrogen. We propose here to extend this to the current situation where suspensions are solidified at a constant cooling rate and starting at room temperature. The following scenario can be applied to the initial instants of solidification and explain the current experimental observations:

- the temperature is progressively lowered from room temperature. When the temperature becomes negative, nucleation of ice crystals can occur.

- until nucleation occurs and nuclei reach their critical size, the system enters a supercooled state; the temperature is lower than the equilibrium temperature, so that it is driven far away from equilibrium.

- when the ice nuclei reach their critical size and keep growing, the system wants to recover its equilibrium, therefore the greater the degree of supercooling, the higher the initial interface velocity. We propose that in this system, nucleation is governed by the presence of impurities, that is, particles in our case. The lower the particle size, the greater the number of available nucleation sites. When plotting the freezing temperature vs. specific surface area of the powders (figure 4), the relationship between both seems relatively straightforward, although more data points would be desirable.

The influence of particle size can now be understood in a relatively straightforward manner. Large particle will offer less nucleation sites, so that the system will be highly supercooled when nucleation finally occurs. The interface velocity vs. time trajectory should be independent of the particle size: the starting velocity is related to the initial degree of supercooling, and hence dependent on particle size, but the following interface displacement kinetics are controlled by the thermal properties of the system, and therefore should not be related to the particle size. The common trajectory observed in figure 1 for all particle size seems to be in good agreement with these explanations. The diffusivity of the particles is of course directly related to the particle size[12], and this could have a great influence on the interface stability, as reported previously[11]. In the initial instants, the interface velocity is nevertheless too high; particles do not have enough time to rearrange or redistribute by diffusion mechanisms.

This suggests a few possibilities to control the initial morphological transition and the resulting structure. The influence of an electric field, driving the particles away or towards the nucleation surface, has already been illustrated[9]. Recent and intriguing results of the influence of the electric field on the freezing temperature of supercooled water have also been reported[13], hinting at a possible influence on the nucleation behaviour, a result that may well be applied to freeze-casting. Another possibility that comes to mind is to start with ice crystals template, an idea already successfully applied[3] to the control of the orientation of the ice crystals and structural features of the freeze-casting process. The initial instants were not investigated in this case, although a great deal of information could be retrieved from such experiments. In regards of the present results, we should expect the initial transition zone to be absent.

## 5. Conclusions

Based on X-ray radiography and tomography experiments of the solidification of alumina suspensions with various particle sizes, under conditions similar to that used in the freeze-casting process, we can draw the following conclusions:

- when cooling starts from room temperature, the suspension enters a supercooled state before nucleation and subsequent crystals growth can occur, driving the system away from equilibrium.
- the nucleation stage is controlled by the particle size, the surface of the ceramic particles acting as nucleation sites. The smaller the particles, the larger the surface area and the higher the number of nucleation sites.
- larger degree of supercooling will lead to faster interface propagation during the initial instants of solidification, which leads to a larger structural gradient in the resulting materials.

**Acknowledgements**
We acknowledge the European Synchrotron Radiation Facility for the provision of synchrotron radiation beam time and we would like to thank E. Boller and J.-P. Valade for their irreplaceable assistance in using beamline ID19. Financial support was provided by the National Research Agency (ANR), project NACRE in the non-thematic BLANC programme, reference BLAN07-2_192446.

# References

[1] C. A. L. Colard, R. A. Cave, N. Grossiord, J. A. Covington, and S. A. F. Bon, "Conducting Nanocomposite Polymer Foams from Ice-Crystal-Templated Assembly of Mixtures of Colloids," *Adv. Mater.,* **21** 1-5 (2009).

[2] S.-W. Yook, H.-E. Kim, and Y.-H. Koh, "Fabrication of porous titanium scaffolds with high compressive strength using camphene-based freeze casting," *Mater. Lett.,* **63**[17] 1502-1504 (2009).

[3] E. Munch, E. Saiz, A. P. Tomsia, and S. Deville, "Architectural control of freeze-cast ceramics through additives and templating," *J. Am. Ceram. Soc.,* **92**[7] 1534-1539 (2009).

[4] S. Deville, E. Maire, A. Lasalle, A. Bogner, C. Gauthier, J. Leloup, and C. Guizard, "In Situ X-Ray Radiography and Tomography Observations of the Solidification of Aqueous Alumina Particles Suspensions. Part I: Initial Instants," *J. Am. Ceram. Soc.,* **92**[11] 2471-2488 (2009).

[5] M. Bettge, H. Niculescu, and P. J. Gielisse, "Engineered porous ceramics using a directional freeze-drying process*,"* pp. 28-34 in *Proceedings of the 28th International Spring Seminar on Electronics Technology.* 2005.

[6] T. Moritz and H.-J. Richter, "Ceramic Bodies with Complex Geometries and Ceramic Shells by Freeze Casting Using Ice as Mold Material," *J. Am. Ceram. Soc.,* **89**[8] 2394-2398 (2006).

[7] T. Moritz and H.-J. Richter, "Ice-mould freeze casting of porous ceramic components," *J. Eur. Ceram. Soc.,* **27**[16] 4595-4601 (2007).

[8] S. W. Sofie, "Fabrication of Functionally Graded and Aligned Porosity in Thin Ceramic Substrates With the Novel Freeze-Tape-Casting Process," *J. Am. Ceram. Soc.,* **90**[7] 2024-2031 (2007).

[9] Z. Yumin, H. Luyang, and H. Jiecai, "Preparation of a Dense/Porous BiLayered Ceramic by Applying an Electric Field During Freeze Casting," *J. Am. Ceram. Soc.,* **92**[8] 1874-1876 (2009).

[10] M. C. Gutierrez, M. Ferrer, and F. del Monte, "Ice-Templated Materials: Sophisticated Structures Exhibiting Enhanced Functionalities Obtained after Unidirectional Freezing and Ice-Segregation-Induced Self-Assembly," *Chem. Mater.,* **20**[3] 634-648 (2008).

[11] S. Deville, E. Maire, G. Bernard-Granger, A. Lasalle, A. Bogner, C. Gauthier, J. Leloup, and C. Guizard, "Metastable and unstable cellular solidification of colloidal suspensions," *Nature Materials,* **8** 966-972 (2009).

[12] S. S. L. Peppin, M. G. Worster, and J. S. Wettlaufer, "Morphological instability in freezing colloidal suspensions," *Proc. R. Soc. London, Ser. A,* **463**[2079] 723-733 (2007).

[13] D. Ehre, E. Lavert, M. Lahav, and I. Lubomirsky, "Water Freezes Differently on Positively and Negatively Charged Surfaces of Pyroelectric Materials," *Science,* **327**[5966] 672-675 (2010).

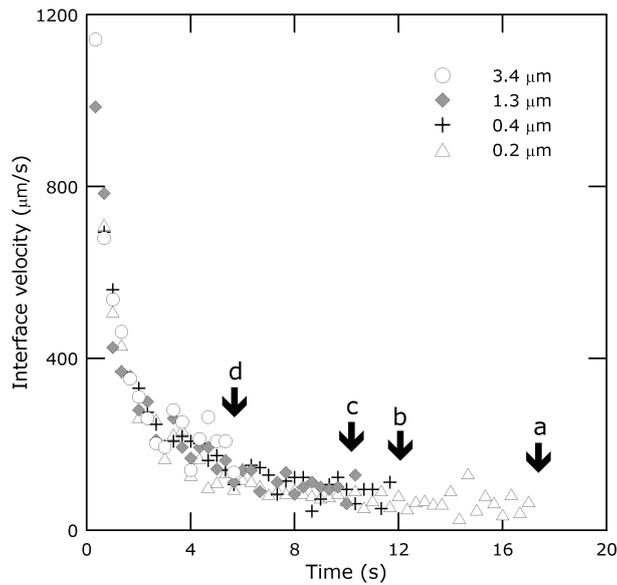

Figure 1: Interface velocity during the initial instants, influence of particle size. The arrows indicate the time at which the interface reaches the top of the observation windows (upper limit of figure 2).

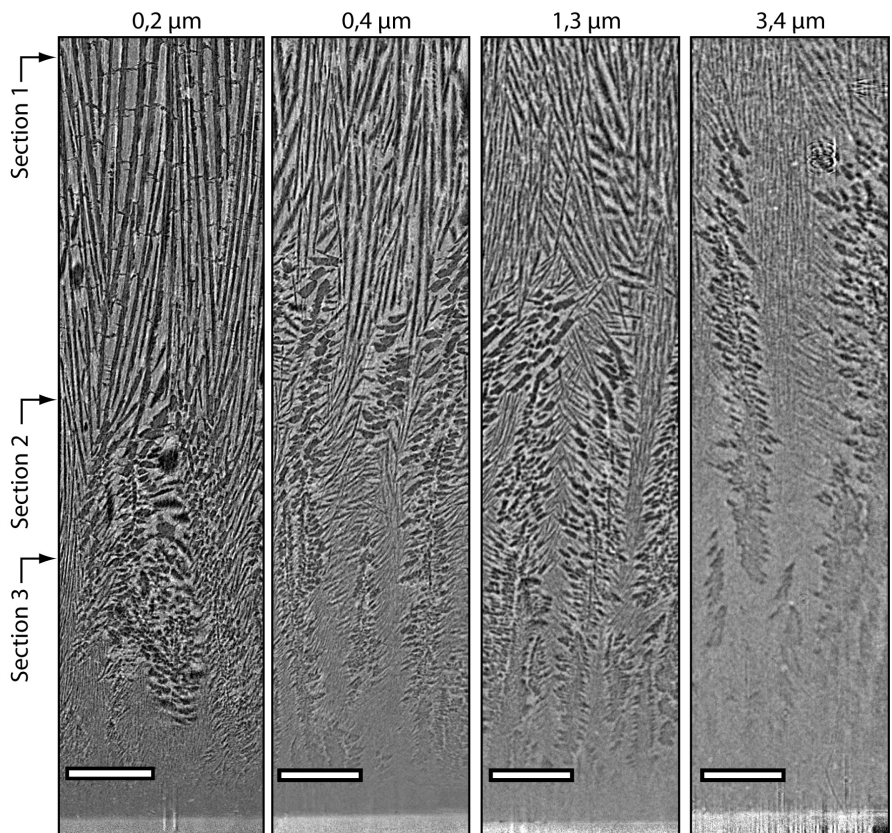

Figure 2: Initial transition zone, influence of particles size. Scale bars 250 µm. Vertical cross-sections. Sections 1-3 refer to cross-sections in figure 3.

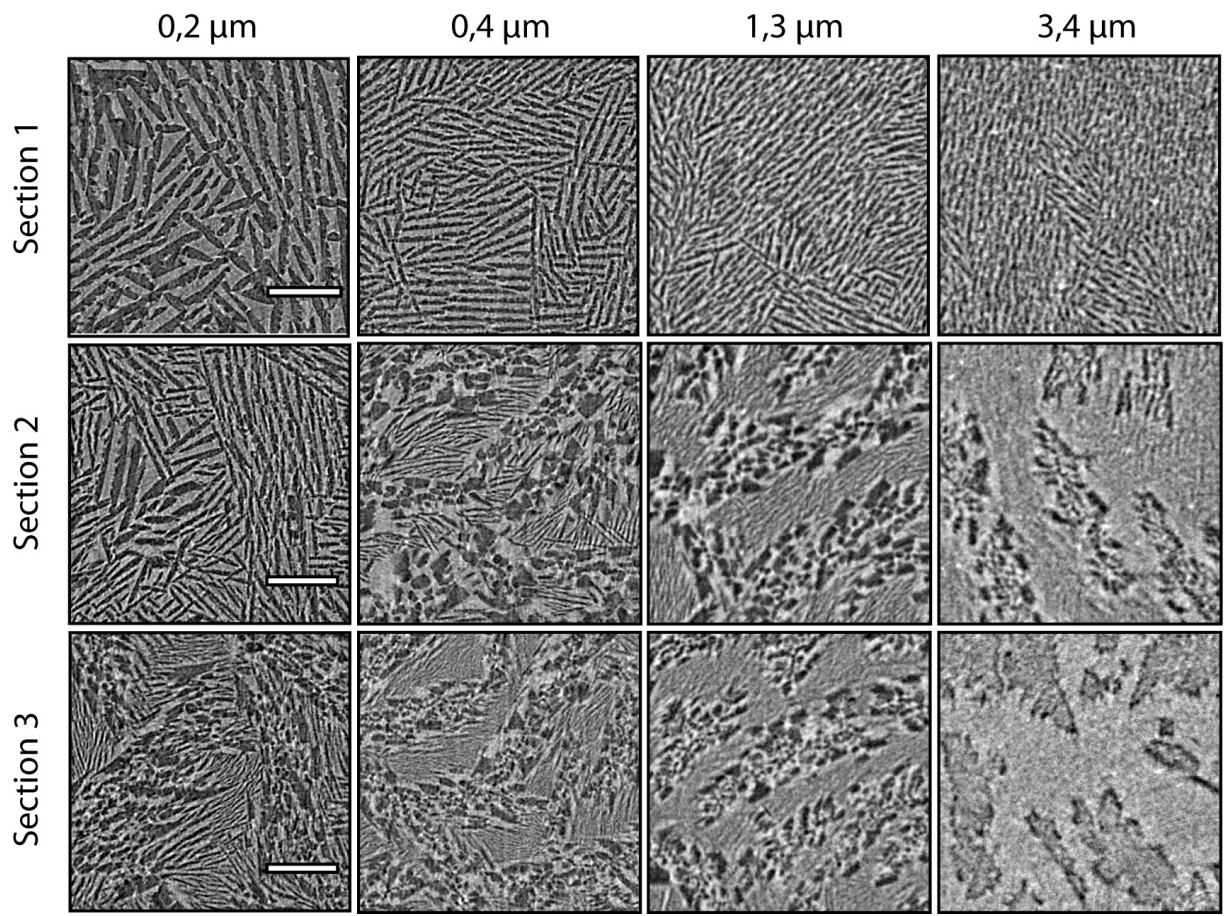

Figure 3: Initial transition zone, influence of particles size. Scale bars 150 µm. Horizontal cross-sections. The different sections are defined in figure 2. The cohabitation of r- and z-crystals (see reference [4] for definition) during the initial instants is clearly visible.

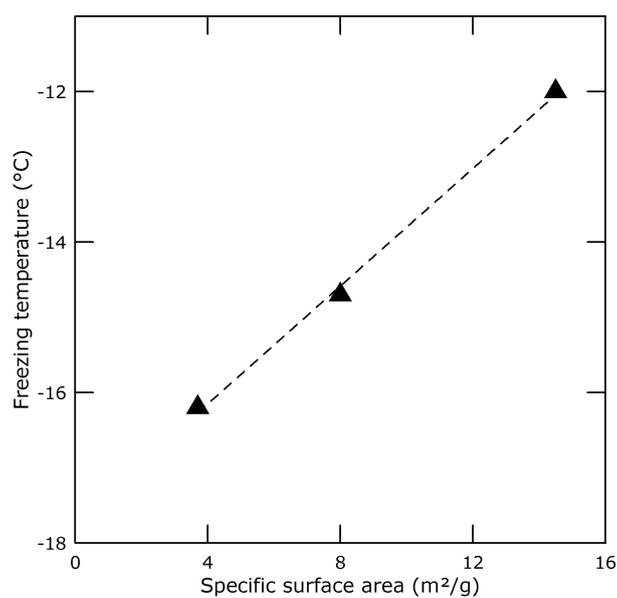

Figure 4: Freezing temperature vs. specific surface area. The nucleation temperature for the 3.4 µm powder is not available.